\documentclass[aps,prb,twocolumn,groupedaddress,showpacs]{revtex4}

\usepackage{epsfig}

\begin{document}


\title{The zero-field superconducting phase transition obscured by finite-size effects in thick
$\mathrm{\bf{YBa_{2}Cu_{3}O_{7-\delta}}}$ films}

\author{M.~C. Sullivan}
\email[Email:]{conroy@physics.umd.edu}
\author{D. R. Strachan}
\author{T. Frederiksen}
\author{R. A. Ott}
\author{M. Lilly}
\author{C. J. Lobb}
\affiliation{Center for Superconductivity Research, Department of
Physics, University of Maryland, College Park, MD 20742}



\begin{abstract}
We report on the normal-superconducting phase transition in thick
$\mathrm{YBa_{2}Cu_{3}O_{7-\delta}}$ films in zero magnetic field.
We find significant finite-size effects at low currents even in
our thickest films ($d = 3200$~\AA).  Using data at higher
currents, we can unambiguously find $T_c$ and $z$, and show $z =
2.1 \pm 0.15$, as expected for the three-dimensional XY model with
diffusive dynamics. The crossover to two-dimensional behavior,
seen by other researchers in thinner films ($d \leq 500$~\AA),
obscures the three-dimensional transition in both zero field and
the vortex-glass transition in field, leading to incorrect values
of $T_c$ (or $T_g$), $\nu$, and $z$.  The finite-size effects,
usually ignored in thick films, are an explanation for the wide
range of critical exponents found in the literature.
\end{abstract}


\maketitle


Since the discovery of high-temperature superconductors and the
realization that their higher critical temperatures and smaller
coherence lengths create an experimentally accessible critical
region \cite{chris}, researchers have looked at these
superconductors in an effort to determine the model that governs
the phase transition. Fisher, Fisher and Huse \cite{ffh} codified
the scaling approach to the normal-superconducting (N-S) phase
transition and predicted the existence of an N-S phase transition
in field, called the vortex-glass transition.  This phase
transition in field has been extensively studied using
current-voltage ($I-V$) isotherms, and although a consensus has
emerged that a vortex-glass transition exists, there is little
consensus in the values of the critical exponents $\nu$ and
$z$\cite{consensus0,consensus1,consensus2}. Moreover, some have
claimed that scaling data collapse does not prove the existence of
a phase transition\cite{against} and that screening can create a
non-zero resistance\cite{screening}, destroying the transition.
Recent work has questioned the existence of a phase transition,
showing that data collapse alone is too flexible, and proposing a
criterion to determine whether or not a phase transition has
occurred\cite{doug}.

In zero magnetic field the existence of an N-S phase transition is
not in doubt.  Very close to $T_c$ ($|T-T_c| \leq 2$
K)\cite{chris}, the transition is expected to obey the
three-dimensional (3D) XY model, with $\nu \approx 0.67$ and $z =
2.0$ for diffusive dynamics \cite{ffh}. Although specific heat and
penetration depth measurements have found mean-field values of
$\nu \approx 0.5$\cite{gauss}, others have fit specific heat and
penetration depth data using critical models with smaller
residuals than for mean-field models\cite{crit}, and recent
thermal expansivity data is more consistent with 3D-XY scaling,
($\nu \approx 0.67$)\cite{thermex}.  Transport measurements can
determine both $\nu$ and $z$, but data in zero field is
inconsistent: Researchers have found vortex-glass like exponents
($\nu = 1.1$, $z = 8.3$) in small fields ($<$ 10
mT)\cite{lowfields}, others finding 3D-XY-like exponents when
extrapolating to zero field from higher fields\cite{moloni} and in
crystals\cite{yeh1}.

Researchers have shown that, in thin films ($d \leq 500$ \AA), the
fluctuation dynamics can cross over from $D=3$ to $D=2$, and that
this crossover occurs at a well-defined current density, $J_{min}$
\cite{finitesizes,ffh}.  In this work we present a systematic
study of $J_{min}$ in films of different thicknesses.  We find
that even in our thickest film ($d=3200$ \AA) the crossover to
$D=2$ obscures the phase transition, causing incorrect choices for
$T_c$, $\nu$, and $z$.  However, at currents greater than
$J_{min}$, we see behavior as predicted by scaling which gives
reliable values for $T_c$ and $z$.

\begin{figure}
\centerline{\epsfig{file=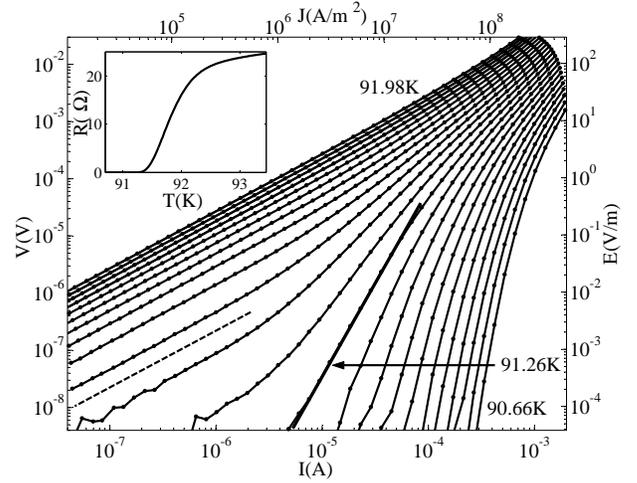,width=3.2in}}
\caption{$I-V$ curves for a 2100
\AA~$\mathrm{YBa_{2}Cu_{3}O_{7-\delta}}$ film, with bridge
dimensions $20 \times 100~ \mu\mathrm{m}^2$, in zero magnetic
field. Isotherms are separated by 60 mK. The dashed line indicates
a slope of 1, or ohmic behavior.  The error bars are smaller than
the points.  The inset is $R(T)$ at 10 $\mu$A. } \label{fig:ivejs}
\end{figure}

We examined the zero-field $I-V$ curves of
$\mathrm{YBa_{2}Cu_{3}O_{7-\delta}}$ (YBCO) films deposited via
pulsed laser deposition onto $\mathrm{SrTiO_3}$ (100) substrates.
X-ray diffraction verified that our films are of predominately
c-axis orientation, and ac susceptibility measurements showed
transition widths $\leq 0.25$ K. $R(T)$ measurements (inset to
Fig.\ \ref{fig:ivejs}) show $T_c \approx 91.5$ K and transition
widths of about 0.7 K.  AFM and SEM images show featureless
surfaces with a roughness of $\approx 12$ nm.  Our films also have
a high critical current ($J_c(77~\mathrm{K}) \approx 2 \times
10^{10} \mathrm{A/m}^2$).  These films are of similar or better
quality than most YBCO films reported in the literature.

Our films were photolithographically patterned into 4-probe
bridges of varying widths (8 - 200 $\mathrm{\mu m}$) and lengths
(40 - 1000 $\mathrm{\mu}$m) and etched with a dilute solution of
phosphoric acid without noticeable degradation of $R(T)$.  We
surround our cryostat with $\mu$-metal shields to reduce the
ambient field to $2 \times 10^{-7}$ T inside. To reduce external
noise, the cryostat is placed inside a screen room with low-pass T
filters at the screen room wall and low-pass $\pi$ filters at the
cold end of the probe.

Figure\ \ref{fig:ivejs} shows $I-V$ curves taken on a 2100
\AA~thick film on a bridge of dimensions $20 \times
100~\mu\mathrm{m}^2$. Scaling predicts\cite{ffh}
\begin{equation}
V \xi^{2+z-D} / I= \chi_{\pm} ( {I\xi^{D-1}}/{T} ),
\label{eq:scaling}
\end{equation}
where $D$ is the dimension, $z$ is the dynamic critical exponent,
$\xi$ is the coherence length, and $\chi_{\pm}$ are the scaling
functions for above and below the transition temperature $T_c$.
Fluctuations are expected to have a typical size $\xi$ which
diverges near $T_c$ as $\xi \sim |T/T_c-1|^{-\nu}$, defining the
static critical exponent $\nu$.

\begin{figure}
\centerline{\epsfig{file=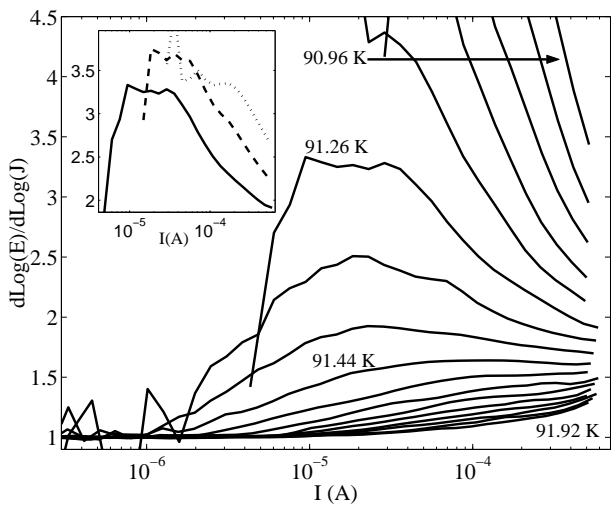,clip=,silent=,width=3.2in}}
\caption{$d$Log$E/d$Log$J$ vs. $I$ for the $I-V$ curves from Fig.\
\ref{fig:ivejs}.  Isotherms are separated by 60 mK.  The
conventional choice for $T_c$, 91.26 K, is clearly not a
horizontal line.  The opposite concavity criterion can be seen at
higher currents ($I > 40$ $\mu$A) about 91.44 K.  The inset shows
the 91.26 K isotherm for three bridge widths on the same film: 20
$\mu$m (solid line), 50 $\mu$m (dashed line) and a 100 $\mu$m
(dotted line), which do not agree as a function of $I$.}
\label{fig:derivi}
\end{figure}

Above $T_c$ at low currents, the $I-V$ curves are expected to be
ohmic (represented in Fig.\ \ref{fig:ivejs} as a dashed line with
slope 1), whereas at higher currents the isotherms are expected to
show non-linear, power law behavior (slope greater than 1).
 Exactly at $T_c$, the coherence length diverges while the
voltage remains finite, which is true only if $V \propto
I^{(z+1)/2}$ (for $D=3$), i.e. a straight line on a log-log plot
\cite{ffh,consensus0}. Conventionally, $T_c$ is chosen as the
first isotherm without an ohmic tail, the isotherm at 91.26 K in
Fig.\ \ref{fig:ivejs}. Data at higher currents and voltages are
typically excluded from fits because it is assumed that the system
is being driven out of thermal equilibrium.  The thick solid line
at 91.26 K is a fit to a power-law at lower voltages, and gives a
dynamic exponent $z = 5.5$, similar to exponents found
elsewhere\cite{lowfields}, but clearly not the expected $z=2$.

We have suggested\cite{doug} that a better way to determine the
critical isotherm is to examine the derivatives of log$E$ vs.
log$J$ isotherms. On such a graph, the critical isotherm would be
obvious as a horizontal line with intercept $(z+1)/2$, separating
isotherms with positive and negative slope (corresponding to
concave up and down in Fig.\ \ref{fig:ivejs}). Our opposite
concavity criterion\cite{doug} states that isotherms at equal
temperatures away from $T_c$ should show opposite concavity at the
same current level.

The derivative plot for the $I-V$ curves in Fig.\ \ref{fig:ivejs}
is shown in Fig.\ \ref{fig:derivi}.  There is no isotherm which is
horizontal over the entire range of currents, contrary to
theoretical expectations. The opposite concavity criterion is also
not satisfied, and isotherms below 91.44 K have unexpected
behavior: they are concave down at higher currents before
displaying ohmic behavior at lower currents. If we consider only
the higher currents ($I > 40 \mu$A), we can see behavior as
predicted by scaling which also satisfies the opposite concavity
criterion: the isotherm at 91.44 K is horizontal, lower isotherms
are concave up, and higher isotherms are concave down.  This
allows an unambiguous choice for $T_c$, $91.44$ K.  If we fit the
high-current data to a horizontal line, then $z = 2.1 \pm 0.15$,
which agrees with diffusive dynamics. Below we will justify
analyzing only $I > 40 \mu$A, ignoring the low-current linear
behavior in these $I-V$ curves (the ohmic ``tails").

\begin{figure}
\centerline{\epsfig{file=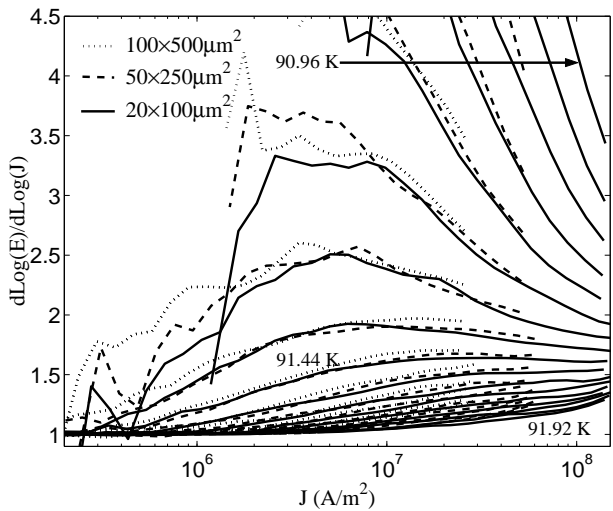,clip=,silent=,width=3.2in}}
\caption{$d$Log$E/d$Log$J$ vs. $J$ for three bridges of different
widths on the same 2100 \AA~film ($20 \times 100
\mu\mathrm{m}^{2}$, $50 \times 250 \mu\mathrm{m}^{2}$, $100 \times
500 \mu\mathrm{m}^{2}$). The crossover to ohmic behavior clearly
depends on $J$.}
 \label{fig:derivj}
\end{figure}

To determine whether the ohmic tails are a bulk intrinsic effect,
we patterned bridges of different widths on the same film.  The
inset to Fig.\ \ref{fig:derivi} shows the 91.26 K isotherm for
three bridges on the same 2100~\AA~film from Figs.\
\ref{fig:ivejs} and \ref{fig:derivi}. Each bridge was measured
simultaneously to insure identical temperatures.  It is clear that
the isotherms do not agree as a function of $I$.  In Figure\
\ref{fig:derivj}, we plot $d$log$E/d$log$J$ as a function of $J$
rather than $I$. All three bridges have similar behavior in $J$,
showing that we are measuring a bulk effect as opposed to an edge
effect\cite{edge}. It is also clear that each isotherm turns over
towards ohmic behavior at a certain applied current density rather
than current. This is significant, because at an applied current
density $J$ one probes fluctuations of typical
size\cite{ffh,finitesizes} $L_J = ( c k_B T / \Phi_{\mathrm{o}} J
)^{1/2}$, where $\Phi_{\mathrm{o}} = h/2e$ is the magnetic flux
quantum and $c$ is a constant expected to be of the same order as
the YBCO anisotropy parameter, $\gamma \approx 0.2$. Thus, as $J$
decreases, $L_J$ increases and will eventually reach the thickness
of the film. At this point a crossover to 2D behavior is expected,
as the size of the fluctuations is limited along the c-axis.
Thus, for a film of thickness $d$, \textit{there is a minimum
current density, such that smaller current densities probe 2D
fluctuations}:
\begin{equation}
 J_{min} = c k_B T / \Phi_{\mathrm{o}} d^2.
\label{eq:jmin}
\end{equation}
Because $J_{min}$ does not depend on the exponents $\nu$ and $z$,
this minimum current density applies for both the vortex-glass
transition and the transition in zero field.

We examined the ohmic tails generated in seven films of different
thicknesses to determine how $J_{min}$ varies as a function of
thickness. We measured seven films with similar properties ($T_c$
and $\Delta T_c$) which varied in thickness from 950 \AA~to 3200
\AA\ \cite{anote}.  To choose $I-V$ curves to compare between
films, we have taken the isotherm which, from high-current data,
most seems like $T_c$, i.e. horizontal on the $d$log$E/d$log$J$
vs. $J$ plot. For $J_{min}$ we have chosen a similar criterion as
Ref.\cite{theysawittoo}, when $d$log$E/d$log$J =
1.2$\cite{anothernote}. If the ohmic tails are caused by
finite-size effects, then we expect $1/\sqrt{J_{min}}$ vs. $d$ to
be a line with slope $(\Phi_{\mathrm{o}}/c k_B T)^{1/2}$, which
will give a value for the undetermined constant $c$. The
temperature $T$ of the different isotherms only varies from 91.4 K
- 92.5 K, a total change of about 1\%.

\begin{figure}
\centerline{\epsfig{file=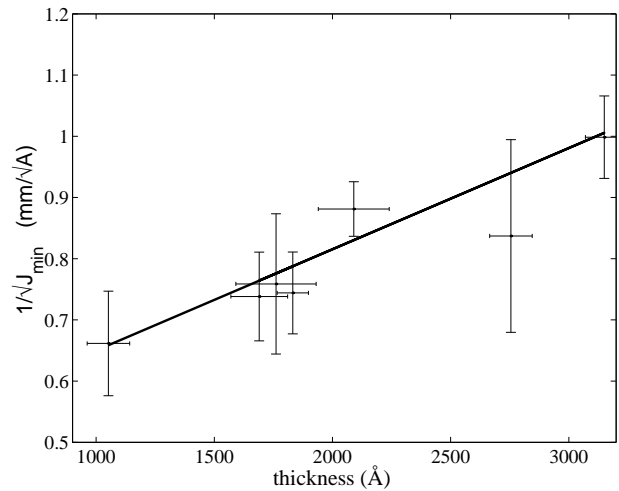,clip=,silent=,width=3.2in}}
\caption{$1/\sqrt{J_{min}}$ vs. thickness for eight different
films.  The solid line is a weighted least-squares linear fit to
the data, with a reduced chi-squared $\tilde{\chi}^2 = 0.41$.  The
slope of the line gives $c=0.60 \pm 0.17$, of the same order as
$\gamma \approx 0.2$, as expected.}
 \label{fig:jmin}
\end{figure}

The results are plotted in Figure \ref{fig:jmin}.  Each value for
$1/\sqrt{J_{min}}$ incorporates error in $I_{min}$, bridge width
and thickness, leading to error bars of about $\pm 22\%$.
Nonetheless, the trend is clear: as $d$ increases, $J_{min}$
decreases.  The solid line in Fig. \ref{fig:jmin} is a weighted
least-squares linear fit to the data with a reduced chi-squared of
$\tilde{\chi}^2 = 0.41$ \cite{fudging}. From the slope we
determine $c = 0.60 \pm 0.17$, the same order of magnitude as
$\gamma \approx 0.2$, as expected.

These finite-size effects have been seen by other researchers in
thinner films.  Dekker \textit{et al.}\cite{theysawittoo} found
$z=2.2 \pm 0.4$ from high-current data in a 500 \AA~thick YBCO
film in zero field, and saw ohmic tails at low currents.  Using an
equation nearly identical to Eq.\ \ref{eq:jmin}, Dekker \textit{et
al.} noted that the fluctuation size along the c-axis saturated at
$\approx$ 470 \AA, as expected.

Finite size effects have also been seen in single crystals, both
in a field and in zero field. Yeh \textit{et al.}\cite{yeh1} found
ohmic deviations from their data collapse at low currents and
attributed them to finite size effects. The length derived from
Eq.\ \ref{eq:jmin} agreed well with the distance between twin
boundaries, and they suggested these boundaries limited the size
of the fluctuations. Although they found good agreement between
theory and experiment, the deviations were determined after the
data collapse. We have shown that $I-V$ curves can be made to
scale with different choices of $T_g$, $\nu$, and $z$\cite{doug},
thus apparent agreement with scaling via a data collapse is not
conclusive evidence that a phase transition occurs, or that one's
choice of critical parameters is the correct one.

W\"{o}ltgens \textit{et al.}\cite{finitesizes} found deviations
from 3D-scaling which appeared as ohmic tails in films with $d
\leq 500$~\AA, as compared to a 3000 \AA~film. W\"{o}ltgens
\textit{et al.} assume that the finite-size effects in thin films
do not extend to the 3000 \AA~film because the $I-V$ curves for
the 3000 \AA~film scale with typical vortex-glass exponents,
despite the fact that Yeh \textit{et al.} found finite size
effects in crystals, where the distance between twin boundaries
were $\approx 2\ \mu$m, nearly an order of magnitude thicker than
W\"{o}ltgens \textit{et al.}'s thickest films. Moreover, a simple
data collapse is not conclusive evidence that the 3000 \AA~thick
films are unaffected by the finite-size effects they see in
thinner films. For $T = 83$ K, $J_{min} \approx 1 \times 10^6
\mathrm{A/m}^2$, and data below $J_{min}$ were included in their
analysis of the 3D transition, and included in the scaling
collapse, possibly affecting their choice of $T_g$.  This
indicates that assuming a data collapse $\textit{a priori}$ and
analyzing deviations from this collapse is not the correct method
to determine finite size effects.

Because the crossover to $D=2$ can affect the choice for $T_c$, it
is an explanation for the wide range of critical exponents found
in the literature. $I-V$ curves are expected to be ohmic at low
currents for $T > T_c$ (or $T_g$), thus it is possible to confuse
ohmic tails generated by finite-size effects with ohmic tails
generated by the 3D phase transition.  This changes the
conventional choice for $T_c$ (or $T_g$) (the first isotherm
without an ohmic tail), and because the ohmic tails are used to
determine $\nu$ (as $R \propto (T/T_c -1)^{\nu(z-1)}$ at low
currents), then values for $\nu$ and $z$ will also be affected.
For example, Sawa \textit{et al.}\cite{japan} scaled $I-V$ curves
for films as thin as 180~\AA~and as thick as 10,000~\AA~in a 2 T
applied field by systematically changing $T_g$, $\nu$ and $z$,
citing this as evidence for the need for an anisotropic 3D-XY
model.  We suggest that the crossover to $D=2$, occurring at
different current densities, required them to vary $\nu$ and $z$
in order to scale their data.

Our results indicate that low-current ohmic tails are due to
finite-size effects.  Failure to account for this leads to
significant underestimates of $T_c$ (or $T_g$) and incorrect
values of $\nu$ and $z$.  We show in the derivative plot that the
high current data agree with the opposite concavity criterion and
lead to unambiguous choices of $T_c$ and $z$.  Because the source
of the low-current ohmic tails is in question, this leaves only
the data collapse to find $\nu$, but using data collapse to find
the critical exponents is perilous \cite{doug}. We can collapse
the data using $T_c$ and $z$ found from the high-current data,
which yields $\nu \approx 1.2$, similar to values found elsewhere
\cite{lowfields,dickhead}, but this value for $\nu$ is clearly not
3D-XY, and other values of $T_c$, $\nu$, and $z$ can collapse the
data also.  It is also unclear how the ohmic tails affect data at
higher currents, especially when $T$ is far from $T_c$ and the
critical region is small.

It is also interesting to note that as $d$ increases, $I_{min}$
does not decrease without limit.  Because $I =  J(wd)$, $I_{min} =
(w/d)(c k_B T/\Phi_o)$, using Eq.\ \ref{eq:jmin}.  The smallest
$I_{min}$ can be is when $w = d$, or $I_{min} \approx 1 \times
10^{-7}$ A ($T = 90$ K). Thus \textit{any} applied current below
0.1 $\mu$A will probe fluctuations limited by the thickness of the
sample, independent of whether the sample is a thin film, thick
film, or single crystal.

In conclusion, we have looked at YBCO microbridges of various
widths (8-200 $\mu$m) in seven films of different thicknesses
(950-3200~\AA) whose zero-field $I-V$ curves are consistent with
low-current ohmic tails created by finite-size effects, even in
the thickest films.  In contrast, the behavior at currents greater
than $J_{min}$ ($I > 40 \mu$A in our film) agrees with the
opposite concavity criterion as predicted by scaling, and gives
the expected 3D-XY dynamic exponent of $z = 2.1 \pm 0.15$. Because
finite-size effects are usually ignored in thicker films, we
suggest that the low-current ohmic tails thought to be the
expected behavior for $T
> T_c$ are actually generated by finite-size effects at
temperatures $T > T_c$ \textit{and} $T < T_c$.  This effect will
obscure the phase transition in all films, both in zero and
non-zero magnetic field, leading to incorrect results for the
critical exponents and temperatures.

The authors thank J.~S. Higgins, A.~J. Berkley, Y. Dagan, M.~M.
Qazilbash, C.~P. Hill, Amlan Biswas, Hamza Balci, R.~A. Headley
and R.~L. Greene for their help and discussions on this work.  We
especially thank N.-C. Yeh for her insights on current-dependent
length scales, and  D. Tobias for suggesting we try bridges of
different widths on the same film. We acknowledge the support of
the National Science Foundation through Grant No. DMR-0302596.


\begin{thebibliography}{99}
\bibitem{anote}
Films thinner than 950~\AA~and thicker than 3500~\AA~tend to have
a lower $T_c$ and a broader $\Delta T_c$.
\bibitem{anothernote}
This choice is arbitrary and other values give similar results.
\bibitem{fudging}
The line fit in Fig. \ref{fig:jmin} does not have a zero
intercept, as one would expect from Eq. \ref{eq:jmin}.  This most
likely results from a low choice of $J_{min}$.  The transition
from a power law to ohmic occurs over more than a decade, and
$d$log$E/d$log$J = 1.2$ is towards the lower end of this
transition. However, the non-zero intercept does not significantly
change the value of $J_{min}$.
\end{thebibliography}
\end{document}